\documentclass[useAMS,usenatbib]{mn2e}

\newcommand\kmsec{~km~s$^{-1}$}

\usepackage{times}
\usepackage{amssymb}
\usepackage{graphicx}
\usepackage{natbib} 

\title[A binary system of tailed radio galaxies]
{A binary system of tailed radio galaxies}
\author[I. Klamer, R. Subrahmanyan and R.W. Hunstead]{I. Klamer$^{1}$\thanks{E-mail:
klamer@physics.usyd.edu.au}, R. Subrahmanyan$^{2}$ and R.W. Hunstead$^{1}$ 
\\\footnotesize$^{1}$School of Physics, University of Sydney, NSW 2006, Australia\\ 
$^{2}$Australia Telescope National Facility, CSIRO, Locked Bag 194, Narrabri NSW 2390, 
Australia}

\begin{document}

\date{}

\pagerange{\pageref{firstpage}--\pageref{lastpage}} \pubyear{2003}

\maketitle

\label{firstpage}

\begin{abstract}
We present a detailed study of a binary system of tailed radio galaxies which, along with 3C75, is the only such binary known to exist. The binary is located in a 
region of low galaxy density at the periphery of a poor cluster Abell S345, but lies close to
the massive Horologium Reticulum supercluster. The radio sources have bent tail morphologies
and show considerable meandering and wiggling along the jets, which are collimated throughout
their lengths. This work presents observations of the large-scale-structure environment of
the binary tailed radio sources with a view to examining the influence
of large-scale flows on the morphology and dynamics of the associated
radio tails. We argue that the orbital motions of the host galaxies together with tidal accelerations
toward the supercluster have resulted in the complex structure seen in these radio tails. 
 
\end{abstract}


\section{Introduction}

Narrow-angle tailed (NAT) radio sources or head-tail (HT) sources 
consist of twin jets issuing from the active nuclei in some galaxies
and then bending by an angle of almost 90\degr. These sources are
usually found in rich cluster environments, where the jets are
understood to have been swept back by the deflecting pressure of the
dense intracluster medium (ICM) arising from the relative motion
between the host galaxy and that medium. 
Studies have shown that in order to bend these tails, 
the ICM must be hot and dense and the relative velocity 
between the radio galaxy and the ICM must be of order 
1000\kmsec \citep{1994ApJ...436...67V}. These required 
velocities and densities make it unlikely for NATs to 
exist in low density environments. 

Wide-angle-tail (WAT) radio sources, on the other hand, 
have beams which typically bend through angles less 
than 90\degr. These sources are, once again, usually found in rich clusters; 
however, the WATs are usually associated with a dominant 
galaxy --- often the central 
galaxy --- which is not expected to be moving at high speeds
with respect to the ambient medium. Therefore, the cause for 
their bending has been unclear: motions of the parent galaxy
with respect to the ICM, buoyancy of the relatively light synchrotron
plasma in the ICM thermal gas, weather/turbulence in the ICM associated with dynamical evolution and cluster-cluster mergers/interactions
may all play roles in particular circumstances.

In this paper we do not attempt to sub-classify the tailed
radio galaxies into WAT and NAT sources because 
the distinction might sometimes be confused by projection effects.
Instead, we refer to them simply as bent-tail (BT) sources.

Comparisons between the radio morphologies of BT sources
in cluster environments 
and the distribution of the intra-cluster thermal gas as seen, for example,
in X-ray images, indicate that the thermal gas is almost always asymmetric
and aligned towards the direction of the bending \citep{gomez}.
It is currently believed that these clusters are undergoing mergers,
resulting in large scale flows of hot gas owing to the changing
gravitational potential. The bending of the BTs is a 
consequence of these flows.  BTs in merging clusters are therefore
a diagnostic of the ICM weather and, consequently, of the evolving
gravitational potential resulting from the merger.

\begin{figure*}
\begin{center}
\includegraphics[width=10cm]{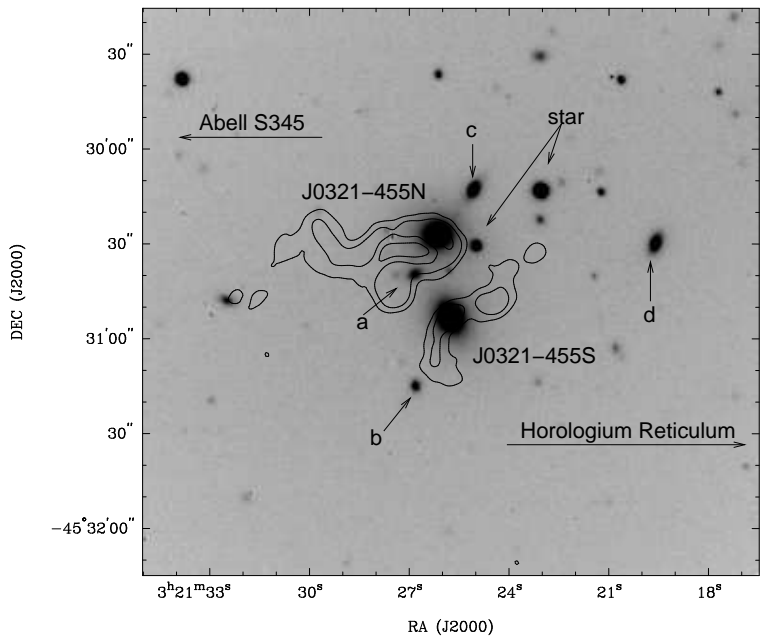}
\caption{\small Binary radio galaxies J0321--455N \& J0321--455S. 
2.4~GHz radio contours are overlaid on a combined R and V band 
optical image of the 
field made using the ANU 2.3-m telescope. The contours correspond 
to $4, 16, 64, 128$ times the rms image noise of $60~\mu$Jy~beam$^{-1}$. 
Spectroscopy (see Table~\ref{2.3mdata}) reveals that the objects 
labelled c and d and the radio galaxies are at about the same redshift.} 
\label{image:CG}
\end{center}
\end{figure*}

Our attention was drawn to the unusual nature of the radio source 
J0321--455 when it appeared in the field of the giant radio galaxy 
PKS~B0319--454 \citep{1994MNRAS.269...37S} with a 
four-leaf-clover structure.  Follow-up observations at higher
resolution revealed the source to have an extremely unusual
configuration: as shown in the overview in Fig.~\ref{image:CG},
J0321--455 is a close pair of twin-tailed 
radio galaxies associated with what appears to be a binary galaxy system.
We label the northern radio galaxy J0321--455N and the southern 
source J0321--455S. The other optical
objects in the field of Fig.~\ref{image:CG}
are labelled alphabetically and discussed in 
Section~2.3. The system  is located in a region of low galaxy density, at 
the periphery of the poor Abell cluster S345 \citep{1989ApJS...70....1A},
and in the vicinity of the massive Horologium-Reticulum supercluster.
The directions toward S345 and the supercluster are indicated in the figure. 

The radio tails in J0321--455 are probably a result of, and a diagnostic for,
(i) the motions of the host ellipticals about their centre of mass, 
and (ii) the relative motion of the binary with respect to the 
intergalactic thermal plasma as a result of 
the gravitational potentials of the large scale structure. 
Therefore, we have undertaken a study of the radio structures and 
the large-scale galaxy environment in this unique situation 
with the aim of understanding the roles of gas and galaxy dynamics in the
triggering and formation of BTs.
In this paper we present radio and optical observations of the binary 
system and its large-scale environment.  Our aim was to image the tailed
radio structures and understand the large-scale mass distribution in order to infer the dynamics and kinematics of the system. The determination of dynamical
and spectral ages are inputs to understanding 
how the nuclear activity in both members of this binary system
might have been triggered and fuelled. The paper is structured 
as follows: Section~\ref{observations} describes the multiwavelength 
observations which include high resolution radio imaging, 
spectroscopy of the nearby objects in Fig.~\ref{image:CG}
and the brightest members of S345, and multifibre spectroscopy 
of a 2\degr field to map the 3-dimensional (3D) large-scale structure. 
Section~\ref{discussion} presents a discussion wherein we have
developed constraints on the source formation based on the
observational data on the radio structures and the large scale environment.  
Throughout the paper we adopt cosmological parameters 
$\Omega_{\Lambda}=0.7$, $\Omega_m=0.3$ 
and $H_0=71$~km s$^{-1}$Mpc$^{-1}$.  S345 and the binary galaxy system are at
a mean redshift of about $z=0.07$ and at this distance,  
1$'$ corresponds to a linear size of 71~kpc.

\section{Observations}
\label{observations}

\subsection{Radio Observations}
\label{radio:obs}
\begin{table*}
\small
\begin{center}
\caption{\small Journal of radio observations made with the ATCA 
and the parameters
of the radio images produced therefrom.}
\begin{tabular}{lcccc}
\hline
 & 1.4~GHz & 2.4~GHz & 4.8~GHz & 8.6~GHz\\
\hline
Observing epoch & 2001 May \& Aug & 2001 May \& Aug  & 1992 Oct & 1992 Oct\\
Array configurations & 6A, 6B & 6A, 6B & 6C & 6C \\
Obs. time (hrs)& 2 $\times$ 12 & 2 $\times$ 12 & 12& 12 \\
Synthesized beam  & $7\farcs7\times5\farcs4$  & $5\farcs0\times4\farcs2$  
& $3\farcs9\times2\farcs6$  & $1\farcs3\times0\farcs8$  \\
& at  P.A. $-1\fdg8$ & at P.A. $-8\fdg8$  
& at P.A. $-15\fdg0$ & at P.A. $-9\fdg5$ \\
Continuum image rms noise ($\mu$Jy~beam$^{-1}$) & 64 & 60 & 68& 60\\
Flux density of J0321$-$455N (mJy) & 111  & 63 & 38 & 4 \\
Flux density of J0321$-$455S (mJy) & 41  & 22 & 13 & 5 \\
\hline
\end{tabular}
\label{table:imageinfo}
\end{center}
\end{table*}

\begin{table*}
\caption{\small Optical properties of the binary galaxies and their 
close neighbours on the sky. Positions and magnitudes are as measured from ANU 2.3-m images and photometry; redshifts
are derived from ANU 2.3-m spectroscopy. The galaxy type
\textit{abs} signifies an absorption line spectrum, \textit{em} an emission line 
spectrum.}
\small
\begin{tabular}{ccccccc}
\hline
Object & Mag & RA  & DEC  & Exposure time for & Redshift (z) & Type\\
  &  (V) & (J2000) &  (J2000) & the spectra ($\times 10^{3}$~s) 
& $\Delta z=0.0001$ & \\
\hline
\vspace{1mm}
J0321$-$455N& 15.79 &03 21 26.2 & --45 30 27 & 8 & 0.0717 & \textit{abs}\\
\vspace{1mm}
J0321$-$455S & 15.59 &03 21 25.8 & --45 30 53 & 12 & 0.0711 & \textit{abs}\\
\vspace{1mm}
a & 18.90 &03 21 26.8 & --45 30 40 & 4 & 0.1814 & \textit{em}\\
\vspace{1mm}
b & 19.26 &03 21 26.8 & --45 31 15 & 6 & 0.1817 & \textit{em}\\
\vspace{1mm}
c & 17.73 &03 21 25.1 & --45 30 13 & 4 & 0.0700 & \textit{abs}\\
\vspace{1mm}
d & 17.84 &03 21 19.6 & --45 30 30 & 4 & 0.0704 & \textit{abs}\\
\hline
\end{tabular}
\label{2.3mdata}
\end{table*}

\begin{figure*}
\begin{center}
\includegraphics[height=11cm]{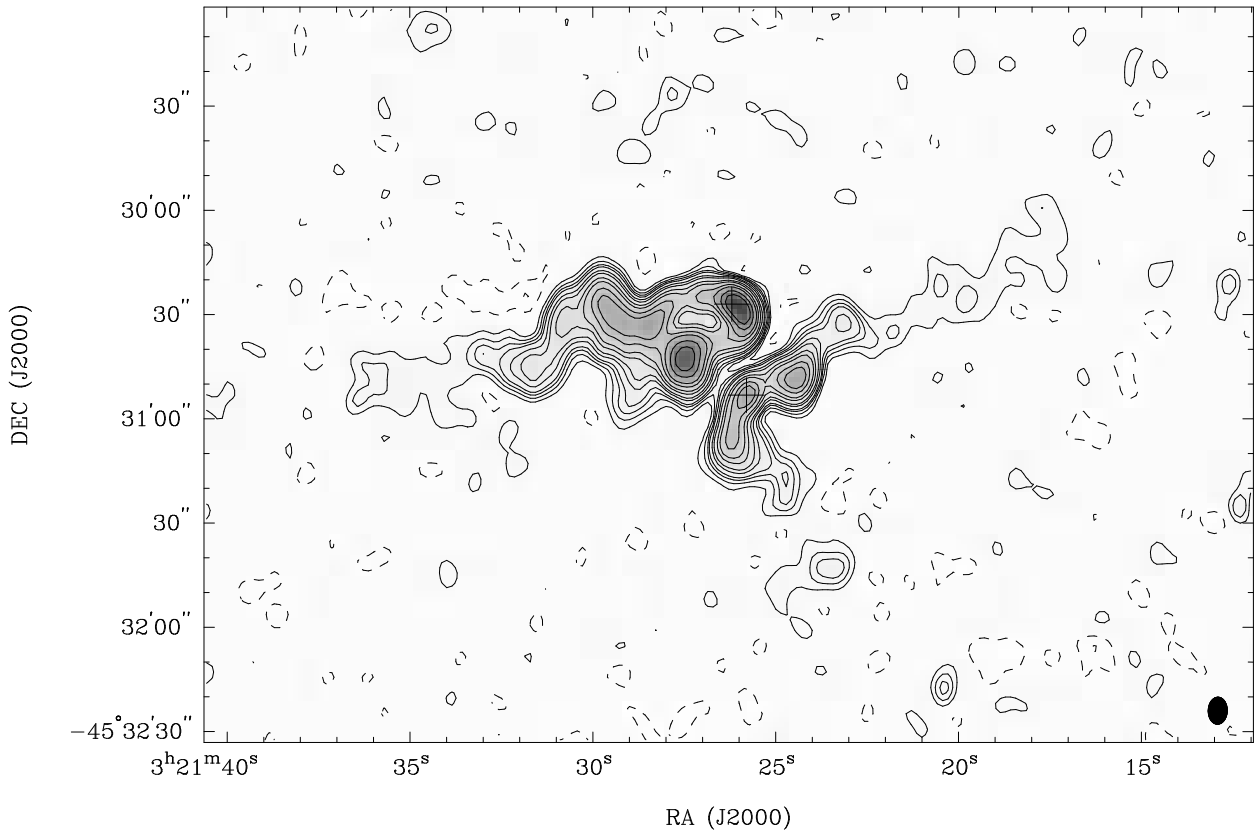}
\caption{\small 1.4~GHz total intensity image.  This image, and the
following radio images, has been corrected for the primary beam attenuation.
Additionally, the positions of the centres of the host optical galaxies are marked 
with crosses and the beam FWHM sizes are shown using filled ellipses at the
bottom right corners.  Contour 
levels are at $\pm1, \pm2, 3, 4, 6, 8, 12, 16, 24, 32, 48, 
64 \times 0.12$~mJy~beam$^{-1}$; the lowest contour is twice the image rms noise.}
\label{image:20cm}
\end{center}
\end{figure*}

\begin{figure*}
\begin{minipage}[l]{7cm}
\centering\includegraphics[height=6cm]{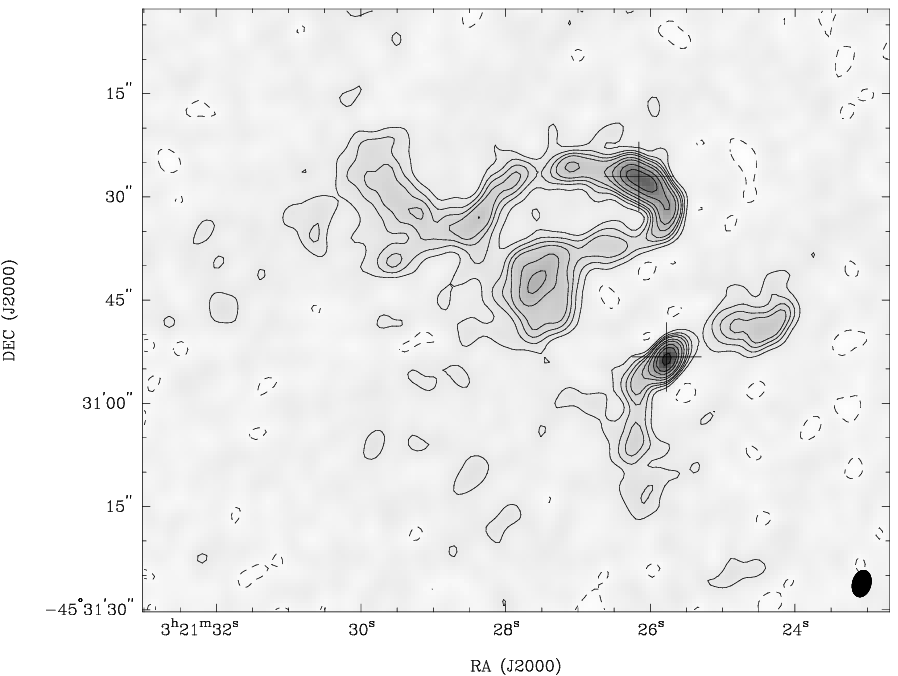}
\caption{\small 4.8~GHz total intensity image. 
Contour levels are shown at  $\pm1, 
2, 3, 4, 6, 8, 12, 16, 24 \times 0.15$~mJy~beam$^{-1}$; the lowest
contour is about twice the image rms noise.}
\label{image:6cm}
\end{minipage} \hspace{1cm}
\begin{minipage}[r]{7cm}
\centering\includegraphics[width=6cm]{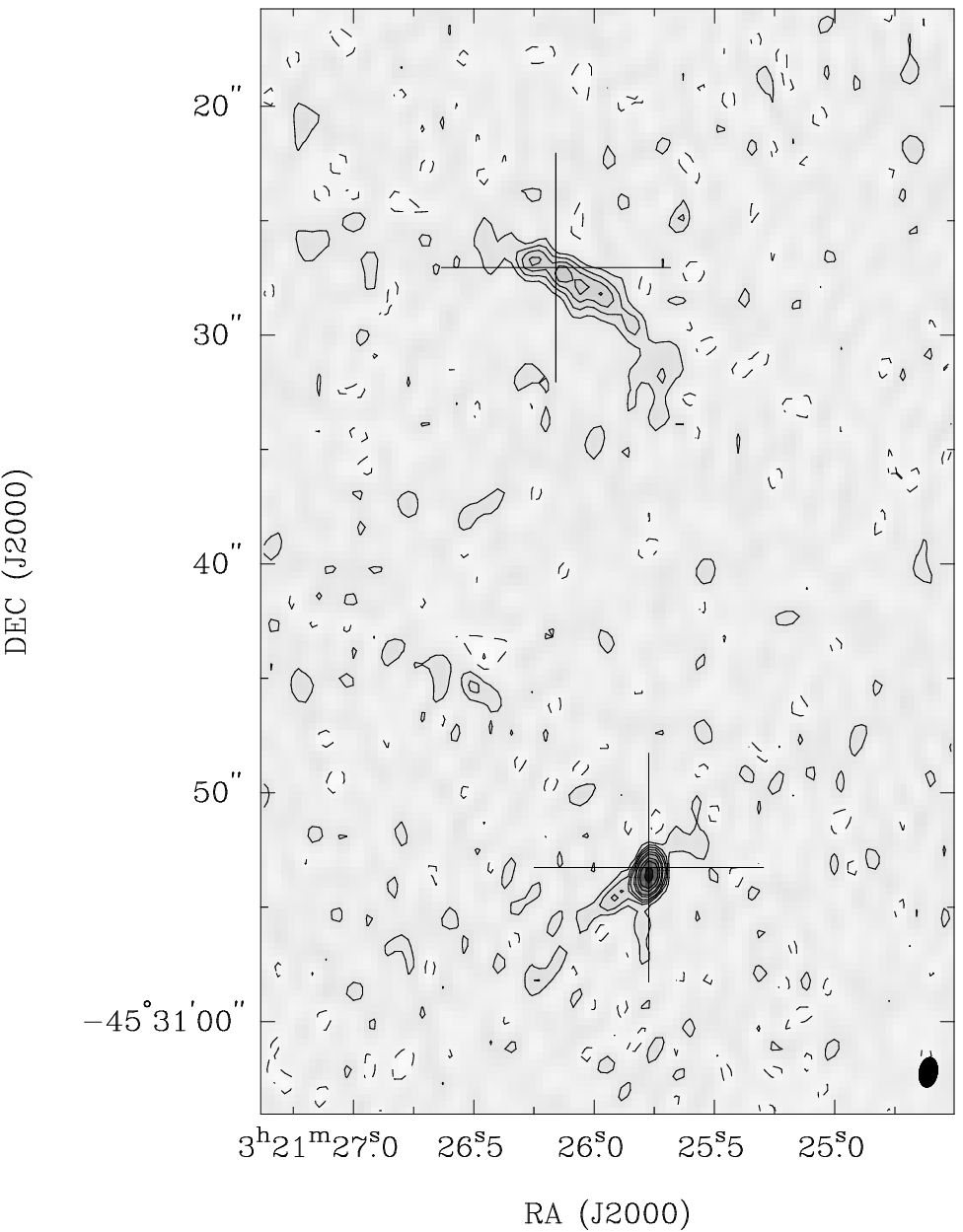}
\caption{\small 8.6~GHz total intensity image. Contour levels are at
$\pm1, \pm2, 3, 4, 6, 8, 12, 16, 24 \times 0.12$~mJy~beam$^{-1}$; the lowest
contour is about twice the image rms noise.}
\label{image:3cm}
\end{minipage}
\end{figure*}

The binary tailed radio sources
J0321--455N and J0321--455S were observed with the Australia Telescope 
Compact Array (ATCA) at 1.4, 2.4, 4.8 and 8.6~GHz; a journal of the radio
observations is given in Table~\ref{table:imageinfo}.  The flux density scale
in all cases was set using observations of PKS~B1934--638; the time
variations in the complex antenna gains as well as the bandpass were
calibrated using frequent observations of a nearby unresolved calibrator
B0332--403.  The visibility data were calibrated and imaged  
using {\textsc{miriad}} following standard ATCA data 
reduction procedures \citep{miriad}.  The beam FWHM sizes and the rms
noise in the final continuum images made at the different observing frequencies
are listed in Table~\ref{table:imageinfo}. The Stokes-I radio images at 
1.4, 4.8 and 8.6~GHz are shown in Figs.~\ref{image:20cm}, \ref{image:6cm} 
and \ref{image:3cm}. All radio images displayed here
and used in the analysis were corrected for the attenuation due to the
telescope primary beam.  The centres of the optical hosts are 
shown marked with crosses in all three images.

The images at the different frequencies and resolutions
show complex structure in the radio tails of both J0321--455N and 
J0321--455S.  The tails of both radio sources appear as collimated jets
that undergo numerous sharp bends and wiggles as they trail away from the
host galaxy.  At 1.4~GHz, the radio tails have been detected out to
projected linear distances of 200~kpc and 150~kpc respectively in 
J0321--455N and J0321--455S.

The surface brightness is a maximum at the locations of the
optical hosts and decreases away from the centres as expected for 
Fanaroff-Riley class I sources (FR~I; \citealp{1974MNRAS.167P..31F}).  
The jets in J0321--455N appear, in projection,
to bend and almost form a closed loop and then
bend sharply in opposite directions; the overall structure in the
inner parts is $\Omega$-shaped as seen clearly in the 4.8~GHz image.
The South-East jet in J0321--455N appears to end in a
plume of higher surface brightness: this structure might simply
be a collimated tail viewed close to the line of sight.
J0321--455S, on the other hand, displays more typical 
WAT morphology.  The jets from this source bend to the same side
and the inner part of this source, as seen in the 4.8~GHz image,
has a C-shaped morphology. A fascinating feature (as seen, for example, in \citealt{hardcastlesakelliou}) of both of these 
bent-jet sources is the sudden drop in surface brightness that occurs
about two-thirds to half the distance to the ends of the tails. 
These lower surface brightness extensions are clearly seen in the
1.4~GHz image but are not detected in the higher frequency images. 
There is considerable meandering in the jet flow
path and fairly abrupt changes in direction in these extensions.
The collimation in the radio jets appears to be maintained 
and there is no evidence for any flaring or transition to wider flows
throughout in spite of the sharp changes in direction. 

In our highest resolution images at 8.6~GHz, all of the extended 
tails fall below the sensitivity of the detectors and the only compact structures detected
at this high frequency are a bright core coincident with the centre
of the galaxy associated with J0321--455S 
and a bright rim representing the inner jets at the centre of
J0321--455N.  The core of J0321--455S is unresolved and has a deconvolved 
size $< 0\farcs3$ and a flux density of 3.76~mJy at 8.6~GHz. 

The total flux density of J0321--455, as computed from the ATCA images,
is 152 and 85~mJy respectively at 1.4 and 2.4~GHz.  The flux densities
are 330 and 230~mJy respectively at 408\footnote{408~MHz flux density measurements based on a re-analysis of the Molonglo Cross archival data by David Crawford (2003, priv. comm.).} and 843~MHz; the PMN survey \citep{PMN}
lists the source flux density as being 60~mJy at 4.85 GHz.  The sources are
at a luminosity distance of 295~Mpc and at 1.4~GHz, 
the radio powers of J0321--455N and J0321--455S are $1.2 \times 10^{24}$ and
$4.3 \times 10^{23}$~W~Hz$^{-1}$ respectively.  These low luminosities are consistent with their FR~I type structure.

\subsubsection{Spectral Index}

The overall spectral index of the source, $\alpha$, is about 0.8 between 843 MHz
and 1.4 GHz (we adopt the definition $S_{\nu} \propto \nu^{-\alpha}$).
However, the total spectrum is flatter at lower frequencies with $\alpha \approx 0.5$ between
408 and 843~MHz.  

To compute the
spectral index distribution between 1.4 \& 2.4~GHz we smoothed the images to a common beam 
of $8\farcs2\times 6\farcs6$ at a position angle of
--6\fdg0.  The spectrum is flattest towards the optical hosts
of the two tailed radio sources; towards the centre of 
J0321--455N $\alpha$ is within the range 0.15--0.25 and towards the centre of 
J0321--455S $\alpha$ is in the range 0--0.1.  J0321--455S has
a compact core that is unresolved in our highest 
resolution observations and the 
spectral index is also the flattest towards this core, implying it may still be active.
Both galaxies also show a steepening in the
two-point spectral index away from their centres and along their radio tails. This trend is consistent
with their FR~I structure.  The spectral 
index takes on values in the range 0.3--1.0 in the higher surface 
brightness tails that are visible in the 4.8~GHz image and steepens 
further to 1.0--2.0 in the
low-surface-brightness extensions that are detected in the 1.4~GHz image.

\subsubsection{Rotation Measure and Polarisation}

We have computed the distribution in rotation measure (RM) over the 
two radio galaxies using the 1.4~GHz and 2.4~GHz position angle images, 
ignoring n$\pi$ ambiguities. We measured the RM for regions whose Stokes~Q and U signal-to-noise ratios were each greater than 2.0 and whose Stokes~I signal-to-noise ratios were greater than 2.5 and where the formal error in position angle was less than 10$^{\circ}$. 

The RM distribution in J0321--455S lies 
within the range 0 to --10~rad~m$^{-2}$ 
(corresponding to a maximum rotation angle of 8$^{\circ}$ at 2.4~GHz) 
over the eastern tail and optical galaxy positions. 
However the RM distribution 
in the local peak in the western tail of J0321--455S varies steadily
across this component by about 40~rad~m$^{-2}$. 
The RM distribution in J0321--455N lies within the range $\pm$20~rad~m$^{-2}$ 
(corresponding to a maximum rotation angle of $\pm$16$^{\circ}$ at 2.4~GHz) 
over the entire structure and, in the vicinity of the optical host, is
observed to vary monotonically through about 40~rad~m$^{-2}$ across the
central component.

The RM due to the Galactic foreground in the direction of these sources 
is estimated to be around $\pm5$ rad~m$^{-2}$.
The fractional polarisation of the radio tails is about 
7--13\% towards the centres (and optical hosts) and rises along the tails, exceeding 60\% at the ends of the 
detected tails.
The projected B field along the northern tail of J0321--455N, and the 
western tail of J0321--455S, appear consistently orthogonal to the 
direction of jet flow, as is usually the case in FR~I radio galaxies. 
The B field appears parallel to the direction of flow in 
the shorter southern jet of J0321--455S and becomes complicated in the 
tight knot of the southern tail of J0321--455N where we believe the 
tail bends into our line-of-sight. 

\subsection{AAT spectroscopy of S345}

Anglo-Australian Telescope (AAT) service spectroscopy of J0321--455N,
J0321--455S and the six brightest members of S345 were obtained on 1999 October 17 using the MITLL2 detector on the RGO spectrograph. Data reduction was performed using routines within the software package \textsc{figaro}. Wavelength calibration was achieved by fitting a fifth order polynomial to short observations of a CuAr comparison lamp. The spectral resolution was 5.4\AA~FHWM corresponding to an equivalent velocity resolution of 300\kmsec at 5500\AA.~Redshifts were measured and we found that the radio galaxies are located close to the mean redshift ($\bar{z}=0.0709\pm0.0005$) of S345.

\subsection{ANU 2.3m Observations}
\label{ANU}

Spectroscopic observations of the sources labelled in Fig.~\ref{image:CG} were 
obtained using the Dual Beam Spectrograph on the Australian National University 
2.3~m telescope at Siding Spring Observatory in 2002 January. The spectra 
were reduced using \textsc{figaro}. Wavelength calibration was computed by fitting a 
third order polynomial, with rms residuals of 0.08\AA, to short exposures of a CuAr comparison lamp. The resolutions of the final spectra were determined to be 5.7\AA~(or 387 km~s$^{-1}$ at 4500\AA) and 
4.3\AA~(or 172 km~s$^{-1}$ at 7500\AA) for the blue and red respectively.  The redshifts 
of each target were determined by cross-correlation with template spectra of 
known redshift using the methods of \citet{1979AJ.....84.1511T}. We used nine 
templates consisting of absorption and emission line galaxy and stellar spectra: these templates were prepared for the 2-degree Field Galaxy Redshift Survey 
\citetext{2dFGRS, \citealp{1998wfsc.conf...77C}}. Regions containing atmospheric emission and absorption lines were systematically 
ignored. The adopted redshift of each source was chosen by selecting the template 
producing the strongest resultant peak in the cross correlation and smallest error. Table~\ref{2.3mdata} 
lists the target sources for the observations and their measured redshifts. Galaxies~c and d 
lie ~340\kmsec~blueward of the radio galaxies, while a and b are 
background galaxies.\newline
\indent R and V images were also obtained on the 2.3~m telescope. Photometric calibration was tied to observations of the Landolt photometric standard star 95-330 and using the \textsc{iraf daophot} analysis package for off-line calibration. The corrected V magnitudes are listed in Table~\ref{2.3mdata}.
\subsection{AAT 2dF Multifibre Spectroscopy}
\label{2df}
A SuperCOSMOS generated catalogue (using the on-line option to separate stars from galaxies) 
shows S345 to be located in a region that is densely populated 
within 1$^{\circ}$ to its West and sparsely populated within 1$^{\circ}$ 
to its East. In order to explore this larger region, multifibre spectra 
with a single pointing centre at (J2000) 03:18:0.0 --45:30:0.0 were obtained 
at the AAT in 2002 January. A magnitude limited (16$<$Bj$<$19.5) 
sample of galaxies was obtained from the SuperCOSMOS catalogue totalling 
660 sources. The sample was reduced to 400 using the B--R colour-magnitude relation for clusters
which uses an estimate of photometric redshifts to reduce
contamination from foreground and background galaxies \citep{2000AJ....120.2148G}. 
The optimal configuration of fibres resulted in 327 allocations, 25 of which 
were used as sky calibrators. The data were reduced using the standard automatic 
2dF data reduction program and spectroscopic redshifts were measured with a 
code written for the 2dFGRS.  Of the 302 target sources, 
13 were misclassified foreground stars and another 13 produced spectra too 
poor to estimate reliable redshifts.   

\subsection{Spectroscopic analysis}
\label{2dfresults}

\begin{figure*}
\begin{center}
\includegraphics[height=8cm]{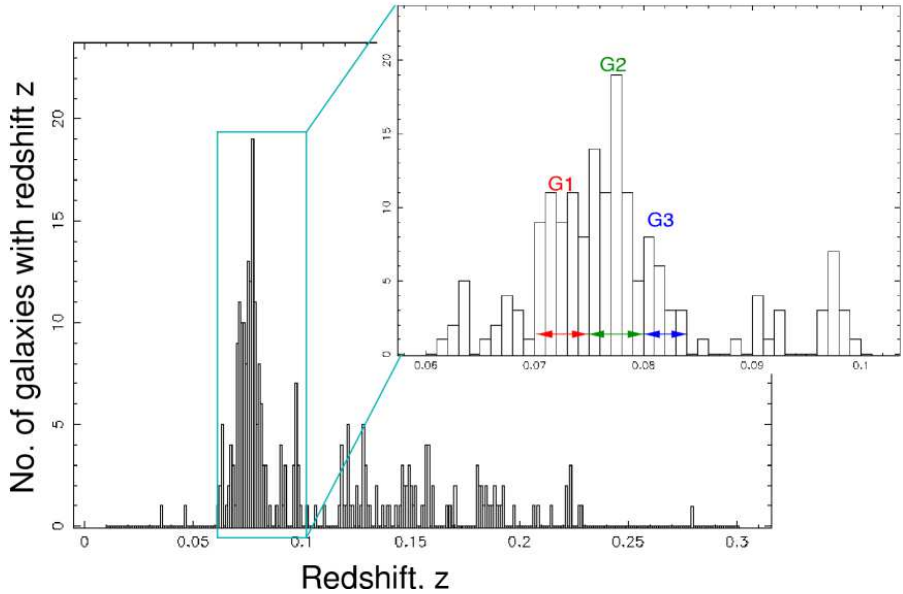}
\caption{\small 2dF redshift distribution histogram. 
Inset shows the trimodal population determined using the KMM algorithm.}
\label{image:histo}
\end{center}
\end{figure*}

The redshift distribution of the remaining 276 galaxies is displayed in the 
histogram shown in Fig.~\ref{image:histo}. We searched the distribution 
for the presence of sub-groupings using the KMM alogorithm~($``$Kaye's$"$ 
mixture model; \citealp{1994AJ....108.2348A}) which fits a user-specified 
number of Gaussian profiles to a data set and assesses the improvement of 
the fit over a single Gaussian distribution. Employing the algorithm over 
the range 0.070$<z<$0.085, and testing for 2-component and 3-component 
Gaussian fits we found the data to be trimodal with 99.8\% confidence. These 
groups, which indicate a distortion of the local Hubble flow, are hereafter 
referred to as G1, G2 and G3 as shown in the inset to Fig.~\ref{image:histo}. 

\begin{figure*}
\begin{center}
\rotatebox{270}{\includegraphics[height=9cm]{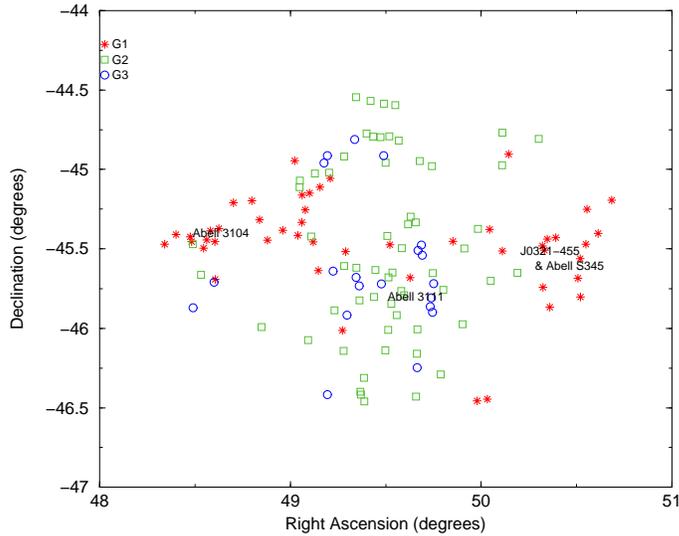}}
\caption{\small Galaxy distribution over the redshift range containing 
galaxy groups G1, G2 and G3 shown in Fig.~\ref{image:histo}. The galaxies are represented as symbols according to their measured redshift; G1 ($0.070\leq z<0.075$) as stars, G2 ($0.075\leq z<0.080$) as squares, and, G3 ($0.080\leq z<0.085$) as circles. Marked are the approximate locations of J0321-455 and several known ACO clusters in the field; refer to Table~\ref{clusterlist} for an exhaustive list including published redshifts.}
\label{image:density}
\end{center}
\end{figure*}

Next, we prepared galaxy distribution plots over the redshift range covering G1, G2 and G3. This distribution is conveyed in Fig.~\ref{image:density} where we plot galaxies as symbols according to their group redshift; galaxies belonging to G1 we plot as stars, G2 as squares and G3 as circles. The most striking feature we see is that G1, which includes the radio galaxies and S345, 
appears to lie along an East-West filamentary 
structure whilst the galaxies within G2 and G3 lie along a plane almost orthogonally to the G1 distribution. For easy reference we also mark the approximate location of the radio galaxies, S345 and a few other known ACO clusters. Known APM \citep{1997MNRAS.289..263D} and ACO clusters in this region are listed in Table~\ref{clusterlist}\footnote{Where a cluster and/or 
group has two identifications, both clusters are listed in Table~\ref{clusterlist}} along with their published redshift and the group, according to our definition, which the cluster resides within. For example, Abell 3111 at $z=0.0775$ resides within G2, whereas Abell 3104 at $z=0.073$ resides within G1. As a side point, Abell 3111 and 3104 are both known members of the Horologium Reticulum (H--R) supercluster.

\begin{figure*}
\begin{center}
\rotatebox{270}{\includegraphics[width=5cm]{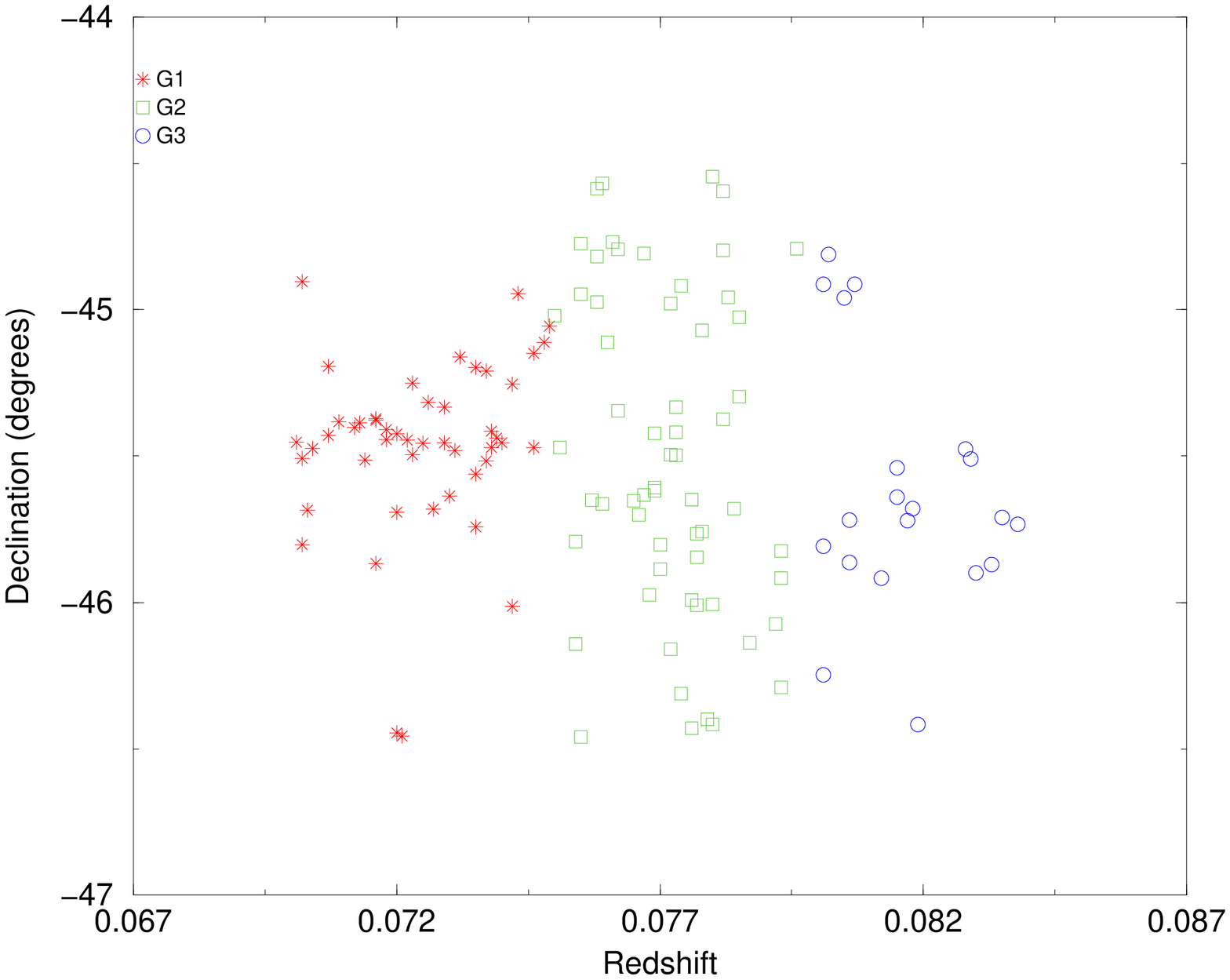}}\hspace{4mm}
\rotatebox{270}{\includegraphics[width=5cm]{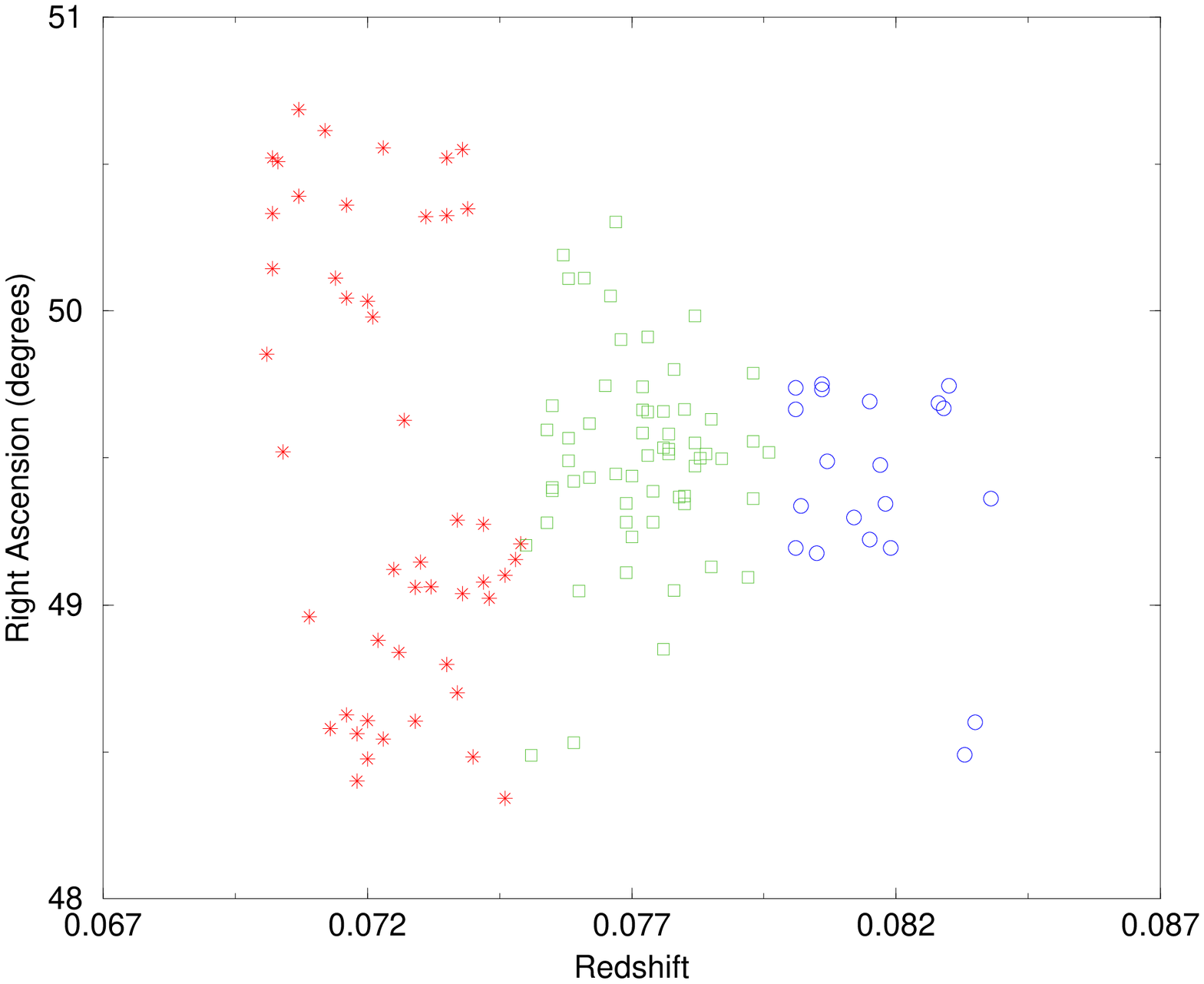}}
\caption{\small Position-redshift plots with established groups G1, G2 and G3 identified.} 
\label{posvelplots}
\end{center}
\end{figure*}

The structure along the North-South axis consists solely of galaxies within G2 and G3 
whereas the galaxy overdensities residing mostly East-West consist almost 
completely of G1 galaxies (which include the radio galaxies). Following the analysis 
of \citet{2002AJ....123.1216R}, a true filament oriented directly North-South should 
be tightly localised in an R.A.$-z$ plot but appear as a vertical line in a 
DEC.$-z$ plot and a filament oriented East-West will show exactly the opposite. 
A candidate group, on the other hand, must be localised in both plots to be physical. 
We prepared position-velocity plots to establish whether the apparent associations 
could be physical entities. These are shown in Fig.~\ref{posvelplots} and clearly 
illustrate that the three groups do appear to be drawn from three separate filamentary 
structures, where G2 and G3 are oriented North-South (and are physically seperate populations) and G1 East-West. 

\begin{table*}
\begin{center}
\caption{\small List of clusters associated with the region surrounding J0321--455 with published redshifts and/or distance classes which fall within G1, G2 or G3 identified in Fig.~\ref{image:histo}}
\small
\begin{tabular}{ccc}
\hline
Cluster & Redshift/Distance Class & Group category\\
\hline
ACO S345 & 0.070 & G1\\
APM 031555-444222 & 0.076  & G2\\ 
ACO S336 & 0.074 & G1 \\ 
APM 366, ACO S334 & 0.075, D$=$4 & G2\\ 
LCLG -45 113, APM 362 & 0.071, 0.072 &G1\\
ACO A3104, LCLG -45 112& 0.073, 0.073&G1\\ 
ACO A3111, APM 367 & 0.0775,0.080 &G2\\
ACO S335  & D$=$4 & G2\\ 
\hline
\end{tabular}
\label{clusterlist}
\end{center}
\end{table*}

\section[]{Discussion}
\label{discussion}
\citet{1997MNRAS.284..286D} defines galaxies located within rich clusters with a velocity difference 
of $\leq$~300~km~s$^{-1}$ and a projected spatial separation of $\leq$70~kpc 
to be binary systems. J0321--455N and 
J0321--455S have a projected separation of 36~kpc and a velocity difference of 
180\kmsec, suggesting that they constitute a gravitationally bound 
binary system. \newline

Notwithstanding the present case, tailed radio sources are extremely uncommon and especially unusual when they come as a pair. In fact, to our knowledge no other binary system of tailed radio galaxies exist (with the possible exception of 3C75). It is therefore strongly suggested that the sources are not independent but that each has been influenced by the presence of the other. Our observational evidence to support this suggestion includes the elongation of J0321--455S along an axis oriented toward J0321--455N and the strong probability that the pair constitute a bound binary system with projected geometry suggestive of clockwise orbital motion. 

\subsection{Connection between the H--R supercluster, S345 and the radio galaxy pair.}

The data presented in Figs.~\ref{image:density} and~\ref{posvelplots} suggest that S345 is part of a filament extending East-West and 
intersecting the dominant cluster A3111. There is another filament containing several known
clusters extending North-South intersecting
A3111. Filamentary structures similar to those we observe have been seen 
in large-scale redshift surveys \citep{2001natur.410} and are 
expected from simulations of large scale structure. 
Since A3111 is a member of the H--R supercluster, 
it is presumably close to the bottom of the local gravitational potential.  
The East-West filament containing S345 and the binary galaxy might be the
path along which galaxies flow (gravitationally) towards the deep local potential well
close to A3111. 
The filamentary nature of the structure containing S345 and the binary galaxy
is probably a consequence of the tidal forces along the filament and we expect that
the binary leads S345 in its infall to the West and towards A3111. The differential
acceleration would imply that the binary currently has a greater velocity to the West
in comparison with S345. The intergalactic medium environment 
would also experience the same tidal acceleration and would be drawn into the
same filamentary structure traced by the galaxies.  In this scenario, the binary might
have been tidally torn from S345 during the infall.

\begin{figure*}
\begin{center}
\includegraphics[width=10cm]{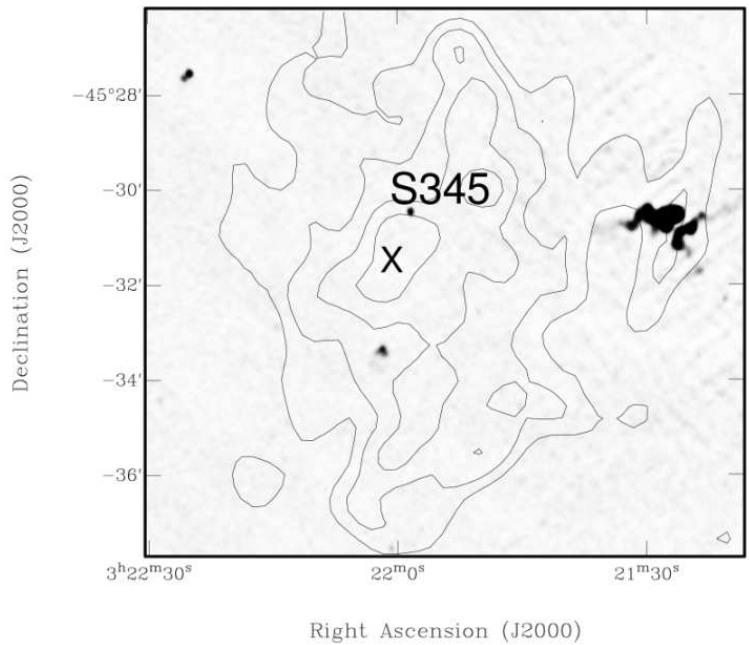}
\caption{\small Smoothed (Gaussian of FWHM 60$''$) ROSAT PSPC X-ray 
contours overlaid on 20cm radio image.} 
\label{image:xray}
\end{center}
\end{figure*}

Fig.~\ref{image:xray} shows the broadband ROSAT PSPC contours (archive image rp800303n00) 
in the region containing S345. The observation was centred on the ACO cluster A3111. 
S345, which lies almost 1\degr from A3111, therefore lies near the edge of the 
PSPC detector's field of view where the point-spread function is significantly distorted. 
Nevertheless, this poor cluster clearly shows X-ray substructure implying that it is
not a relaxed system.  More significant is the detection of X-ray gas associated with the
binary system and the suggestion of a bridge between the gas associated with S345 and the
binary. The structure in the gas is consistent with our interpretation that the binary
has been tidally detached from S345 and, along with its gaseous environment, leads
S345 in its westward accelerated motion.
  
\subsection{What triggered the radio emission?}
Galaxy-galaxy interactions play a vital role in providing fuel to the central black hole of an elliptical galaxy through a nuclear starburst. This starburst can produce enough cold gas to increase the accretion rates surrounding a
black hole and trigger nuclear activity \citep{1990ApJ...350...89B,2000ApJ...545L..93L,2003ApJ...582..668B}. The presence of close neighbours might, therefore, trigger a starburst and initiate/restart nuclear
activity from J0321--455. An obvious interpretation for this system is therefore that both radio galaxies were triggered as a result of the mutual interaction between the massive ellipticals as they travel on close orbits and hence the AGN have a common triggering time. This is consistent with our observations which show the spectral indices along the tails of the pair of galaxies are quite similar.\newline 

Despite the natural conclusion that the radio emission was triggered by galaxy-galaxy interactions, we stress that there are much higher density systems where the incidence of powerful emission is almost non-existent. For example, the high densities in Hickson compact groups (we refer the reader to the review by \citealp{1997ARA&A..35..357H}) make them ideal regions to search for signs of galaxy interactions. Studies of such groups would therefore be expected to show large numbers of powerful radio sources. Interestingly, they do not; see, for example, \citet{1999ApJ...523...87T} who studied radio emission in compact groups with a median redshift of $0.09$ and a distribution of 1.4~GHz luminosities in the range $8\times10^{21}-1\times10^{25}$W~Hz$^{-1}$ with a median of $2\times10^{23}$W~Hz$^{-1}$. The study found that although radio emission was detected in 20\% of compact groups, only 1\% of groups contained a potential tailed radio galaxy and none contained a double tailed source even though several galaxies have luminosities greater than J0321--455. Furthermore, one group (ShCG219) containing an FR~II type radio galaxy shows evidence for strong interaction with another nearby galaxy of similar magnitude and yet this second galaxy shows no sign of having associated radio emission. These results are hard to account for if the triggering of an AGN occurs simply through galaxy-galaxy interactions. 
  
\subsection{Characterising the motion of the galaxies}

Optical imaging and spectroscopy have given us two spatial and one velocity dimension for the host galaxies. Radio imaging has added extra information from which we can infer the transverse motions of the galaxies. We can therefore, albeit crudely, estimate the inclination of the orbital plane to the line of sight. \newline
Under the assumption that J0321--455 is a binary system, the small disparity in their spectroscopically measured redshifts is then due to a line-of-sight velocity difference of 180\kmsec where the sources are moving about each other in a clockwise (East through North) direction such that J0321--455S appears blueshifted and J0321--455N redshifted from their common systemic redshift. Assuming the galaxies have equal masses (as determined by their very similar optical magnitudes) then we infer J0321--455N \& J0321--455S to have line-of-sight velocities of 90\kmsec and --90\kmsec respectively. 

If the line-of-sight velocity is of order the orbital velocity of the binary system, then we can infer an orbital period of $8\times10^8$~years. Since the two radio sources are neither intertwined nor display significant curvature to the locus of their tails, this orbital period should be much longer than the radiative age of the synchrotron particles at the edges of the tails. This is consistent with \citet{2000AJ....119.1111B} who show that the maximum radiative age of a synchrotron emitting particle is a few $\times10^7$~years. 

A crude alternative to infer the orbital velocity of the binary is to assume that the synchrotron plasma of the radio jets are deposited with no bulk kinetic energy. Then we can assume that the extent of the radio tails represents the host galaxy motion through the IGM. Under this assumption, we infer a minimum orbital velocity of $3000$\kmsec~if the synchrotron electrons are as old as $5\times10^7$ years. The inconsistency between these two velocities forces us to conclude that host galaxy motion cannot be the only contributing factor to the morphology, and extent, of the radio tails. In other words, the morphology of the radio tails must be a combined effect of orbital motion plus another force accelerating them East-West relative to North-South.\newline  

The light synchrotron plasma tails of the BT sources are embedded in the denser
thermal environment.  The dynamical evolution of the filament and, consequently, the
tidal expansion of the gaseous environment of the BT sources, might be expected to
lengthen the structures East-West relative to North-South and 
might be the cause of the East-West extensions in the tails from the binary system. 
In this scenario, the extended western jet of J0321--455S lies ahead of the binary 
system and therefore has a relatively greater acceleration toward the supercluster 
than its host galaxy. Similarly, the radio tails of J0321--455N extend behind the 
binary relative to the supercluster and are accelerated more slowly than 
the binary system.  In this interpretation, the morphology of the BT sources is partly consistent with tidal expansion of the environment along an East-West direction due to two influences: the proximity (3~Mpc in projection) of the massive ($5\times10^{15}-5\times10^{16}~M_{\sun}$) supercluster and the less massive but much closer S345 ($10^{13}~M_{\sun}$ located 0.4~Mpc in projection). A factor of two differential expansion of the intergalactic medium in East-West relative to North-South would then require about $2\times10^8$~years.

\subsection{What bends the tails?}

There are several factors that might contribute to bending of the tails of 
BT radio galaxies, and these will probably depend on the environment 
the sources are located within. Some studies have shown that the bending of 
BT sources in poor clusters, with typical observed densities of 
$10^{-4}$~cm$^{-3}$, can be produced if the host galaxy is travelling with velocities of order 
1000 km~s$^{-1}$~\citep{1994ApJ...436...67V,1995AJ....110...46D} 
relative to the ICM gas. This usually means the cluster is virialised, the 
ICM gas is in hydrostatic
equilibrium within the cluster potential, and the galaxies are moving in
the cluster potential and relative to the gas. 
An alternate scenario is one in which the ICM gas has 
streaming flows driven by changing gravitational potentials in cluster
mergers and, as a consequence, the ICM gas may
have an associated bulk velocity relative to the potential and the galaxies. 
In both these cases, the bending is inferred to be due to the deflecting pressure
of the ICM, arising from the relative motion of the galaxy and the ICM. 
With orbital velocities of order 100\kmsec, we reject this possibilility for the morphology of J0321-455.

On the other hand, a wind would surely bend the tails of both galaxies in the same direction, 
as seen in other examples of clusters containing several BTs (e.g. the Abell 
clusters A3266 and A119). The radio images of J0321--455 show that the tails 
in the two tailed radio sources trail off in very different directions.  
Obviously, this cannot be due to relative motion of the binary as a whole 
with respect to the ICM.  

The ROSAT PSPC broadband image detects thermal X-ray gas in the vicinity of the binary and we infer that, if associated with S345 and the radio galaxies, this gas
might have a temperature of about 0.6--1.3 keV \citep{bird} depending
on whether the gas is associated with the binary or the cluster potential.  The radio tails
have a fairly constant width along their lengths, suggesting static confinement
by the ICM thermal gas pressure. Assuming minimum energy conditions \citep{miley}, 
the synchrotron tails have internal pressures about $9 \times 10^{-13}$~dyne~cm$^{-2}$ 
indicating an ambient density of $6 \times 10^{-4}$~cm$^{-3}$. Models  for deflecting 
jets in low luminosity sources suggest that the low relative velocity between the 
host galaxies and the ambient gas, together with the density we infer for the gas, 
are insufficient to cause the bending via deflecting pressures.

Another factor which may play a role in the observed bending of BTs in clusters 
is the existence of turbulence in the intra-cluster gas caused by cluster 
merger-induced shocks. In this case the cluster is no longer virialised and the 
gas becomes heated and turbulent, consistent with observations which show that WATs 
are preferentially located in clusters containing significantly higher levels of 
X-ray emission than a similar sample of radio-quiet counterparts \citep{1998MNRAS.301..609B}. 
There are several reasons why we consider that merger-induced ``cluster weather'' 
might play a role in the case of J0321--455.  First, the results from 
Section~\ref{2dfresults} indicate that Abell S345 and the binary probably have 
a peculiar velocity associated with a gravitational infall into the 
Horologium-Reticulum supercluster. Secondly, there is X-ray evidence suggesting 
Abell S345 is not a relaxed cluster, but rather is in a dynamically
evolving state. Thirdly, S345 and the binary are part of a filamentary structure.

In conclusion, the model we propose is one in which the thermal gaseous environment statically confines the light synchrotron plasma and provides a dense environment (substantiated by the X-ray profile of the group and the collimated radio tails) within which the light radio tails are embedded. The orbital motions
of the binary with respect to the relatively static ambient medium, as well as the 
tidal differential expansion in the ambient gas, have together resulted in the BT morphologies.
The tails, therefore, represent the trails of the moving host galaxies within
the differentially expanding environment. The detailed paths of the tails would
probably be additionally influenced by any turbulence or  
`cluster weather' as a consequence of the evolution in the large scale 
structure, particularly the differential gravitational acceleration towards
the nearby supercluster.

\section{Summary}
\label{summary}
This work has presented a shining example of the influence of large scale flows on the morphology of radio galaxies. We have presented multiwavelength observations of a pair of tailed 
radio galaxies located on the periphery of the poor Abell cluster S345, 
and residing approximately 3~Mpc in projection from the Horologium-Reticulum Supercluster. This system is extraordinary because it is (with the possible exception 3C75) the only known example of a binary system of powerful radio galaxies. Although the obvious triggering mechanism for the pair is their mutual interaction, the lack of other examples in the literature is hard to reconcile with such a statement. We suggest that turbulence and tidal forces due to the nearby cluster and supercluster may be another consideration. 
Optical spectroscopy of the host galaxies and their neighbours on the sky
show that the hosts constitute a binary system.
High-resolution radio imaging has shown that the tails of both galaxies are 
extended East-West along a galaxy filament linking S345 to the supercluster.  
Multi-object spectroscopy of galaxies within a 1\degr~radius shows that S345 
and other nearby clusters have a 3D spatial distribution consistent with 
S345 and the binary being part of a filament that is 
suffering tidal acceleration towards the nearby Horologium-Reticulum supercluster. 
The existence of X-ray substructure indicates a lack of dynamical equilibrium
and is support for turbulent `cluster weather' in the gaseous environment.
We propose that the BT source structure has evolved primarily as a result 
of the combined effects of binary orbital motion and tidal differential expansion in the environment
over a period less than $10^8$~years. 
\section*{Acknowledgments}

These observations would not have been possible without the help of Russell Cannon (AAO) for the 2dF observations and suggestions on the galaxy distribution analysis and the members of the 2dFGRS Team, led by Will Sutherland, whose runz software allowed us to measure the redshifts of the 2dF galaxies. We are grateful to Matthew Colless for the nine template spectra used for cross-correlation. We thank Lakshmi Saripalli and Ann Burgess for their contributions made early on in this work. We also thank our referee Chris Simpson for his valuable comments. The Australia Telescope is funded by the Commonwealth of Australia for operation as a National Facility managed by CSIRO. This research has made use NASA's Astrophysics Data System and of the NASA/IPAC Extragalactic Database (NED) which is operated by the Jet Propulsion Laboratory, California Institute of Technology, under contract with the National Aeronautics and Space Administration.

\small

\bibliography{mnemonic,mnemonic-simple,mnbib}

\begin{thebibliography}{}

\bibitem[\protect\citeauthoryear{{Abell}, {Corwin} \& {Olowin}}{{Abell}
  et~al.}{1989}]{1989ApJS...70....1A}
{Abell} G.~O.,  {Corwin} H.~G.,    {Olowin} R.~P.,  1989, ApJS, 70, 1

\bibitem[\protect\citeauthoryear{{Ashman}, {Bird} \& {Zepf}}{{Ashman}
  et~al.}{1994}]{1994AJ....108.2348A}
{Ashman} K.~A.,  {Bird} C.~M.,    {Zepf} S.~E.,  1994, AJ, 108, 2348

\bibitem[\protect\citeauthoryear{{Barton Gillespie}, {Geller} \&
  {Kenyon}}{{Barton Gillespie} et~al.}{2003}]{2003ApJ...582..668B}
{Barton Gillespie} E.,  {Geller} M.~J.,    {Kenyon} S.~J.,  2003, ApJ, 582, 668

\bibitem[\protect\citeauthoryear{{Bird}, {Mushotsky} \& {Metzler}}{{Bird}
  et~al.}{1995}]{bird}
{Bird} C.~M.,  {Mushotsky} R.~F.,    {Metzler} C.~A.,  1995, ApJ, 453, 40

\bibitem[\protect\citeauthoryear{{Bliton}, {Rizza}, {Burns}, {Owen} \&
  {Ledlow}}{{Bliton} et~al.}{1998}]{1998MNRAS.301..609B}
{Bliton} M.,  {Rizza} E.,  {Burns} J.~O.,  {Owen} F.~N.,    {Ledlow} M.~J.,
  1998, MNRAS, 301, 609

\bibitem[\protect\citeauthoryear{{Blundell} \& {Rawlings}}{{Blundell} \&
  {Rawlings}}{2000}]{2000AJ....119.1111B}
{Blundell} K.~M.,  {Rawlings} S.,  2000, AJ, 119, 1111

\bibitem[\protect\citeauthoryear{{Byrd} \& {Valtonen}}{{Byrd} \&
  {Valtonen}}{1990}]{1990ApJ...350...89B}
{Byrd} G.,  {Valtonen} M.,  1990, ApJ, 350, 89

\bibitem[\protect\citeauthoryear{{Colless}}{{Colless}}{1998}]{1998wfsc.conf...%
77C}
{Colless} M.,  1998, in Wide Field Surveys in Cosmology, Pub. Editions
  Frontieres {Early Results from the 2dF Galaxy Redshift Survey}.
p.~77

\bibitem[\protect\citeauthoryear{{Dalton}, {Maddox}, {Sutherland} \&
  {Efstathiou}}{{Dalton} et~al.}{1997}]{1997MNRAS.289..263D}
{Dalton} G.~B.,  {Maddox} S.~J.,  {Sutherland} W.~J.,    {Efstathiou} G.,
  1997, MNRAS, 289, 263

\bibitem[\protect\citeauthoryear{{den Hartog}}{{den
  Hartog}}{1997}]{1997MNRAS.284..286D}
{den Hartog} R.,  1997, MNRAS, 284, 286

\bibitem[\protect\citeauthoryear{{Doe}, {Ledlow}, {Burns} \& {White}}{{Doe}
  et~al.}{1995}]{1995AJ....110...46D}
{Doe} S.~M.,  {Ledlow} M.~J.,  {Burns} J.~O.,    {White} R.~A.,  1995, AJ, 110,
  46

\bibitem[\protect\citeauthoryear{{Fanaroff} \& {Riley}}{{Fanaroff} \&
  {Riley}}{1974}]{1974MNRAS.167P..31F}
{Fanaroff} B.~L.,  {Riley} J.~M.,  1974, MNRAS, 167, 31P

\bibitem[\protect\citeauthoryear{{Gladders} \& {Yee}}{{Gladders} \&
  {Yee}}{2000}]{2000AJ....120.2148G}
{Gladders} M.~D.,  {Yee} H.~K.~C.,  2000, AJ, 120, 2148

\bibitem[\protect\citeauthoryear{{Gomez}, {Pinkney}, {Burns}, {Wang}, {Owen} \&
  {Voges}}{{Gomez} et~al.}{1997}]{gomez}
{Gomez} P.~L.,  {Pinkney} J.,  {Burns} J.~O.,  {Wang} Q.,  {Owen} F.~N.,
  {Voges} W.,  1997, ApJ, 474, 580

\bibitem[\protect\citeauthoryear{{Griffith}, {Wright}, {Burke} \&
  {Ekers}}{{Griffith} et~al.}{1994}]{PMN}
{Griffith} M.~R.,  {Wright} A.~E.,  {Burke} B.~F.,    {Ekers} R.~D.,  1994,
  ApJS, 90, 179

\bibitem[\protect\citeauthoryear{{Hardcastle} \& {Sakelliou}}{{Hardcastle} \&
  {Sakelliou}}{2003}]{hardcastlesakelliou}
{Hardcastle} M.~J.,  {Sakelliou} I., , 2003, astro-ph/0312245

\bibitem[\protect\citeauthoryear{{Hickson}}{{Hickson}}{1997}]{1997ARA&A..35..3%
57H}
{Hickson} P.,  1997, ARA\&A, 35, 357

\bibitem[\protect\citeauthoryear{{Lim}, {Leon}, {Combes} \&
  {Dinh-V-Trung}}{{Lim} et~al.}{2000}]{2000ApJ...545L..93L}
{Lim} J.,  {Leon} S.,  {Combes} F.,    {Dinh-V-Trung} 2000, ApJL, 545, L93

\bibitem[\protect\citeauthoryear{{Miley}}{{Miley}}{1980}]{miley}
{Miley} G.,  1980, ARA\&A, 18, 165

\bibitem[\protect\citeauthoryear{{Peacock}, {Cole}, {Norberg} \&
  {Baugh}}{{Peacock} et~al.}{2001}]{2001natur.410}
{Peacock} J.~A.,  {Cole} S.,  {Norberg} P.,    {Baugh} C.~M. e.~a.,  2001, Nat,
  410, 169

\bibitem[\protect\citeauthoryear{{Rose}, {Gaba}, {Christiansen}, {Davis},
  {Caldwell}, {Hunstead} \& {Johnston-Hollitt}}{{Rose}
  et~al.}{2002}]{2002AJ....123.1216R}
{Rose} J.~A.,  {Gaba} A.~E.,  {Christiansen} W.~A.,  {Davis} D.~S.,  {Caldwell}
  N.,  {Hunstead} R.~W.,    {Johnston-Hollitt} M.,  2002, AJ, 123, 1216

\bibitem[\protect\citeauthoryear{{Saripalli}, {Subrahmanyan} \&
  {Hunstead}}{{Saripalli} et~al.}{1994}]{1994MNRAS.269...37S}
{Saripalli} L.,  {Subrahmanyan} R.,    {Hunstead} R.~W.,  1994, MNRAS, 269, 37

\bibitem[\protect\citeauthoryear{{Sault} \& {Killeen}}{{Sault} \&
  {Killeen}}{1999}]{miriad}
{Sault} R.~J.,  {Killeen} N.~E.~B.,  1999, Miriad User's Guide.
http://www.atnf.csiro.au/computing/software/miriad

\bibitem[\protect\citeauthoryear{{Tonry} \& {Davis}}{{Tonry} \&
  {Davis}}{1979}]{1979AJ.....84.1511T}
{Tonry} J.,  {Davis} M.,  1979, AJ, 84, 1511

\bibitem[\protect\citeauthoryear{{Tovmassian}, {Chavushyan}, {Verkhodanov} \&
  {Tiersch}}{{Tovmassian} et~al.}{1999}]{1999ApJ...523...87T}
{Tovmassian} H.~M.,  {Chavushyan} V.~H.,  {Verkhodanov} O.~V.,    {Tiersch} H.,
   1999, ApJ, 523, 87

\bibitem[\protect\citeauthoryear{{Venkatesan}, {Batuski}, {Hanisch} \&
  {Burns}}{{Venkatesan} et~al.}{1994}]{1994ApJ...436...67V}
{Venkatesan} T.~C.~A.,  {Batuski} D.~J.,  {Hanisch} R.~J.,    {Burns} J.~O.,
  1994, ApJ, 436, 67

\end{thebibliography}

\bibliographystyle{mn2e}

\bsp

\label{lastpage}

\end{document}